\documentclass{ieeeaccess}
\usepackage{cite}
\usepackage{amsmath,amssymb,amsfonts}
\usepackage{algorithmic}
\usepackage{graphicx}
\usepackage{textcomp}
\usepackage{float}
\usepackage{caption,setspace}

\DeclareUnicodeCharacter{2212}{-}
\DeclareUnicodeCharacter{2212}{/}

\def\BibTeX{{\rm B\kern-.05em{\sc i\kern-.025em b}\kern-.08em
    T\kern-.1667em\lower.7ex\hbox{E}\kern-.125emX}}

\begin{document}

\history{Date of publication xxxx 00, 0000, date of current version xxxx 00, 0000.}
\doi{}

\title{A Comprehensive Review on the NILM Algorithms for Energy Disaggregation}

\author{\uppercase{Akriti Verma}\authorrefmark{1}, 
\uppercase{Adnan Anwar\authorrefmark{2}}, M. A. PARVEZ MAHMUD\authorrefmark{3}, \uppercase{Mohiuddin Ahmed\authorrefmark{4}}, and \uppercase{Abbas Kouzani\authorrefmark{3}}}

\address[1]{School of Information Technology, Deakin University, Geelong, Australia}
\address[2]{Strategic Centre for Cyber Security Research and Innovation (CSRI), School of IT, Deakin University, Australia}
\address[3]{School of Engineering, Deakin University}
\address[4]{School of Science, Edith Cowan University, Perth, Australia}


\markboth
{Author \headeretal: Preparation of Papers for IEEE TRANSACTIONS and JOURNALS}
{Author \headeretal: Preparation of Papers for IEEE TRANSACTIONS and JOURNALS}

\corresp{Corresponding author: Adnan Anwar (e-mail: adnan.anwar@deakin.edu.au)}

\begin{abstract}
The housing structures have changed with urbanization and the growth due to the construction of high-rise buildings all around the world requires end-use appliance energy conservation and management in real time. This shift also came along with smart-meters which enabled the estimation of appliance-specific power consumption from the building’s aggregate power consumption reading. Non-intrusive load monitoring (NILM) or energy disaggregation is aimed at separating the household energy measured at the aggregate level into constituent appliances. Over the years, signal processing and machine learning algorithms have been combined to achieve this. Incredible research and publications have been conducted on energy disaggregation, non-intrusive load monitoring, home energy management and appliance classification. There exists an API, NILMTK, a reproducible benchmark algorithm for the same. Many other approaches to perform energy disaggregation have been adapted such as deep neural network architectures and big data approach for household energy dis aggregation. This paper provides a survey of the effective NILM system frameworks and review the performance of the benchmark algorithms in a comprehensive manner. This paper also summarizes the wide application scope and the effectiveness of the algorithmic performance on three publicly available data sets. 
\end{abstract}


\titlepgskip=-15pt
\maketitle

\section{Introduction}
\label{sec:introduction}
There has been immense research on developing technological solutions in order to address the energy requirements which are rising exponentially and thereby making the challenge of energy conservation harder day by day. The increasing energy demands not only affects a country's economy but also comes with significant negative implications on the environment. Hence, the only effective way to conserve energy now is to encourage its efficient usage. A fine-grained monitoring of energy consumption and communicating the same to the consumers can help in the noteworthy reduction of energy wastage   ~\cite{shorrock2003domestic},   ~\cite{darby2006effectiveness}. The method for the disaggregation of electrical measurements was proposed by Hart \cite{192069} at the beginning of the 1980s and owing to the shift towards smart meters, ease of data availability in the last few years and amidst rising interests in climate change led to the boost of extensive research in this field. Energy disaggregation primarily aims to provide a piece of detailed information about energy sensing and a breakdown of the consumption patterns, which can further be analysed for the automation of energy management and the development of similar systems. Strategies such as rescheduling the appliances with high-power consumption rates to off-peak times will thus be feasible \cite{zoha2012non}.
It would also help appliance design companies acquire a more reliable understanding of the power-usage patterns of their devices and hence will enable them to develop sustainable products.
There are two primary approaches to energy disaggregation, in particular, Intrusive Load Monitoring (ILM) and Non-Intrusive Load Monitoring (NILM).  Intrusive load monitoring consists of using a low-end metering device for the measurement of the electricity consumption of one or more appliances, usually employing at least one censoring device per appliance, whereas in the case of NILM it only requires a single meter per house or a building that is being monitored. Non-intrusive load monitoring (NILM) or energy disaggregation is aimed at separating the household energy measured at the aggregate level into constituent appliances. The presence of nilmtk-contrib    ~\cite{10.1145/3360322.3360844}, an open-source, implementation of the energy disaggregation problem, has unfolded the means for comparisons of the different algorithms executing energy disaggregation. It has also enabled researchers to observe and evaluate NILM approaches for abstraction and varied evocation as they can be applied to multiple data sets accessible online. The versatility of NILM API makes experimentation in this field easy by lowering the entry barriers and making the implementation generic irrespective of the datsets and algorithms which helped to progress research in this area.

\begin{table*}[]
\footnotesize
\caption{Articles included in systematic literature review}
\begin{tabular}{|p{4cm}|p{4cm}|p{0.6cm}|p{0.6cm}|p{0.6cm}|p{0.6cm}|p{0.6cm}|p{0.6cm}|p{0.6cm}|}

\hline
Subject &
  Purpose &
  Dia-logue on framework &
  Dis-cussion on NILM-API &
  Related dataset references &
  Algo-rithm implementation &
  NILM-API implementation &
  Empir-ical analysis &
  Future research prosp-ects \\ \hline 
Non-Intrusive Load Monitoring System   Framework and Load Disaggregation Algorithms: A Survey\cite{8861646} &
  presents a general NILM framework and reviews publicly available data sets. &
  Yes &
  Yes &
  Yes &
  No &
  No &
  No &
  Yes \\ \hline
A Survey on Non-Intrusive Load Monitoring   Methodies and Techniques for Energy Disaggregation Problem\cite{faustine2017survey} &
  presents an overview of NILM, related methods and reviews state-of-the-art NILM algorithms.&
  Yes &
  Yes &
  Yes &
  No &
  No &
  No &
  No \\ \hline 
Building power consumption datasets:   Survey, taxonomy and future directions\cite{HIMEUR2020110404} &
  analyses the nature of building energy consumption datasets. &
  Yes &
  No &
  Yes &
  No &
  No &
  No &
  Yes \\ \hline
On performance evaluation and machine   learning approaches in non-intrusive load monitoring\cite{klemenjak2018performance} &
  determines the accuracy and generalisation abilities of NILM algorithms using the data sets REDD, UK-DALE, and Dataport. &
  Yes &
  Yes &
  Yes &
  No &
  No &
  No &
  Yes \\ \hline
Prospects   of Appliance Level Load Monitoring in     Off-the-Shelf Energy Monitors: A Technical Review\cite{haq2018prospects} &
 encourages the incorporation of appliance-level e-monitoring and load disaggregation; sets requirements to implement load disaggregation in next-generation e-monitors. &
  Yes &
  No &
  Yes &
  No &
  No &
  No &
  Yes \\ \hline
An Overview of Non-Intrusive Load   Monitoring: Approaches, Business Applications, and Challenges\cite{zhuang2018overview} &
  surveys NILM framework and load disaggregation algorithms; reviews load signature models, exhibits existing datasets and performance metrics. &
  Yes &
  Yes &
  Yes &
  Yes &
  No &
  Yes &
  Yes \\ \hline
  Literature Review of Power Disaggregation\cite{jiang2011literature} &
  reviews the current state of the algorithms and
NILM systems. &
  Yes &
  No &
  No &
  No &
  No &
  No &
  Yes \\ \hline
  Load Disaggregation Technologies: Real World and Laboratory
Performance\cite{mayhorn2016load} &
   reviews recent field studies and laboratory tests of NILM technologies. &
  Yes &
  No &
  No &
  Yes &
  No &
  Yes &
  Yes \\ \hline
\textbf{Proposed~Comprehensive~review on energy disaggregation}&
   \textbf{reviews~of~the state-of-the-art~techniques, implements NILMTK, presents analysis and draws conclusions}&
  \textbf{Yes} &
  \textbf{Yes} &
   \textbf{Yes} &
   \textbf{Yes} &
   \textbf{Yes} &
   \textbf{Yes} &
   \textbf{Yes} \\ \hline

\end{tabular}
\label{Table: LR}
\end{table*}

\section{Scope and contribution of this paper}
There has been a good number of research and publications in energy disaggregation, non-intrusive load monitoring, home energy management and appliance classification. Recent discoveries in machine learning and computer vision has driven a lot of attraction to machine learning-based strategies for NILM.  As it develops, energy consumption datasets serve as the basis for train and test data for  NILM algorithms. These data sets are generated by using a piece of measurement equipment such as a smart plug and combing the energy readings given by smart-meters and appliance-level meters. The increasing popularity of research in this field has led to a large number of datasets being released in the past few years. NILMTK was designed specifically as an open-source implementation to unfold the means for comparisons and proliferation of the different algorithms executing energy disaggregation.    ~\cite{10.1145/3360322.3360844}. Performance evaluation, contrast and comparison of different NILM algorithms is still a research challenge, for multiple reasons    ~\cite{klemenjak2018performance}   \cite{pereira2018performance} \cite{herrero2017non}. Table-\ref{Table: LR} is a tabular depiction of the literature review which integrates the information about the purpose of research and their major findings in this field, that has been compiled from similar survey papers. It can be observed that the prospects of NILM application are both, wide and versatile. There have been multiple attempts to devise the generalisation abilities of existing NILM algorithms in order to incorporate the methodology to appliance-level monitoring and disaggregation. It is also challenging to collect and store data at the required sample rates in order to make data sets available for these algorithms. This paper provides a concise and clear overview of the NILM framework and the NILMTK-API by implementing the API on three publicly available data sets and comparing the observed results.

This survey article exhibits the following concepts:
\begin{enumerate}
    \item This paper aims to present an updated overview of the techniques used in the NILM framework including the recent developments in its strategies for the energy disaggregation problem.
    \item The paper presents an experimental overview of the application of NILM-API, which was released with nilmtk-contrib, on three publicly available data sets, draws conclusions and highlights on future research directions.
    \item A detailed overview of the energy disaggregation problem is presented. Here, we have shown the advantages of the NILM API which enables comparison of algorithms in this domain even without expertise in the field. 
\end{enumerate}

\section{Organisation of this paper}
The rest of the paper is organised as follows. We start with a brief background on the load monitoring approaches which is followed by a formal introduction to the energy disaggregation problem and a discussion about the available implementations and their general framework.  We then introduce the State-of-the-art Disaggregation Techniques, the NILMTK-contrib repository, and summarise the different methods that are being employed for NILM. Next, we describe the benchmark data sets, including the features taken into consideration for the experiment API. Later we, describe the advantages of this implementation by the means of an empirical comparison, before we sum up our analysis.

\section{Background on Load Monitoring approaches}
Load monitoring and identification is a mechanism for the evaluation and regulation of the usage of electrical energy and operating conditions of individual appliances, wherein the main power metre located in the house forms the basis of the analysis of the composite load measurement. They can supply the customer and the utility with information such as the type of load, the detail of the electricity consumption and the operating conditions of the appliances. The practice of load monitoring is quintessential for the management of energy as it gives us statistical insights on appliance energy consumption and their patterns which can be applied for scheduling heavy loads and optimal utilisation of energy resources. 

\subsection{Intrusive load monitoring}
Intrusive load monitoring consists of using a low-end metering device for the measurement of the electricity consumption of one or more appliances.  It is said to be 'intrusive' in order to indicate that the meter which is being used to employ energy monitoring is located inside the dwelling, typically close to the appliance that is monitored. In essence, an intrusive load monitoring system standardises a mechanism for measuring the energy usage of an appliance by employing a metering system that connects each appliance to power meters within the household. It requires, therefore, entering the house and is thus remarked as an intrusive system. The results obtained with intrusive load monitoring are accurate, however, it requires an elaborate installation incurring high costs which usually demands wiring and units for data storage to the households concerned   ~\cite{aladesanmi2015overview}.  The techniques for load monitoring can also be divided into direct or indirect monitoring techniques   ~\cite{6486814}. Direct monitoring techniques are often referred to as physically intrusive procedures, measure the electrical properties of the power of each appliance requirement. The process involves a tiny low-end device that is connected to the power cord of each device, to generate physically intrusive signatures for measuring the energy consumed by the appliances. The state of operation of an appliance is signalled to the data collecting device, whenever the appliance is switched on. The electromagnetic field produced by the current flow through the wire is used to calculate the power drawn by the appliance. Although it provides an accurate measurement of energy consumption, it is not always economically feasible   ~\cite{6486814}.

\subsection{Non-intrusive load monitoring}Non-intrusive load monitoring consists of measuring electricity consumption using a smart meter, which is usually placed at the meter panel. 
It is also known as one-sensor metering since the whole process relies on a single point of measurement.  This method qualifies to be non-intrusive meaning that no extra equipment is installed inside the house. Since this is a one-sensor metering procedure, the signatures of appliances get superposed in order to comprehend the contribution of single appliances, they're needed to be separated. This operation of separating the energy drawn by each appliance is called the disaggregation of the total electricity consumption.  Non-intrusive load monitoring is a simple and reliable way to determine energy usage and therefore the status of operation of specific appliances, carrying the analysis of the aggregate load estimated by the building's main meter. The process of NILM also involves the analysis of individual appliance energy consumption, the deduction of the devices being utilised in a building as well as the changes happening within the voltage and current entering into a building ~\cite{aladesanmi2015overview}. It is considered non-intrusive since it does not involve an intrusion into the building or consumer premises when measuring the energy consumption of various appliances. Smart metres equipped with NILM technology are used by companies to survey the particular uses of electrical power in different homes. NILM is taken into account for the cost-effective alternative to intrusive monitoring techniques. The ideation behind analysing the flow of power in order to record the operating conditions of appliances in a household was coined by George W. Hart while he attempted to collect and examine residential load data as a part of a residential photo-voltaic system   ~\cite{192069}. The acknowledgement of step-change in either active or reactive power out of the total load is the working principle for load monitoring. It is executed by modifying the operating state of the various customer’s appliances. A NILM is installed temporarily to analyse the characteristics of the appliances which may be used to recommend ways for reducing energy demand and costs.

\section{The Energy Disaggregation Problem}
Speeding urbanisation has led to some conventional changes in the housing structures which engrossed the need for the development of high-rise buildings all around the world.  The maintenance of these buildings requires end-use appliance energy conservation in a synchronous manner. The transformation in housing patterns also came together with a mechanism for an appliance level energy break down of the building’s aggregate power consumption. The building's total energy reading can now be used for the estimation of appliance-specific power consumption which is facilitated by the use of smart meters. Nearly, two decades ago,  a method for the disaggregation of electrical measurements was proposed by Hart, which required only the examination of the overall load data in order to identify signatures of the power consumption of each appliance among them. The proposed methodology did not require the installation of any equipment inside the property of the customer and is thus considered to be non-intrusive. The aggregated data for the building's energy usage can be obtained from outside the building or the residence using the main electrical panel. The separation of the whole-house building data into its principal energy-consumption constituents is the complete goal of this process. NILM algorithms have the ability to determine device-specific energy disaggregation using only the aggregated data collected from a single measurement point and thus Non-Intrusive Load Monitoring remains an attractive method for energy disaggregation.

\subsection{General Framework of NILM}
As mentioned in the nilm toolkit  \cite{10.1145/3360322.3360844}  \cite{10.1145/2602044.2602051}, the model on which NILM works is as follows: For an observed time series of aggregate measurements $Y = (Y_1, Y_2,..., Y_T )$, where $Y_t \in R^+$ represents the active power which is measured in Watt-hours or Watts using an electricity meter at time $t$. This signal represents the sum of the total energy consumed by all the constituent appliances present in a building.  For the same scenario, the building facility is assumed to have $I$ appliances, wherein this equation $X_i = (x_{i1}, x_{i2}, ... , x_{iT} )$ where $x_{it} \in R^+$ is the energy-signal description for each appliance. 
According to the disaggregation algorithm, the main readings of the buildings can be expressed as the sum total of the appliance signals along with $E_t$ where $E_t$ is an error term. Thus, when given the aggregate measurement of $Y$, the goal of the energy disaggregation problem is to retrieve the unknown signals $X_i$.

\begin{figure}[]

    \includegraphics[width=0.75\columnwidth,keepaspectratio]{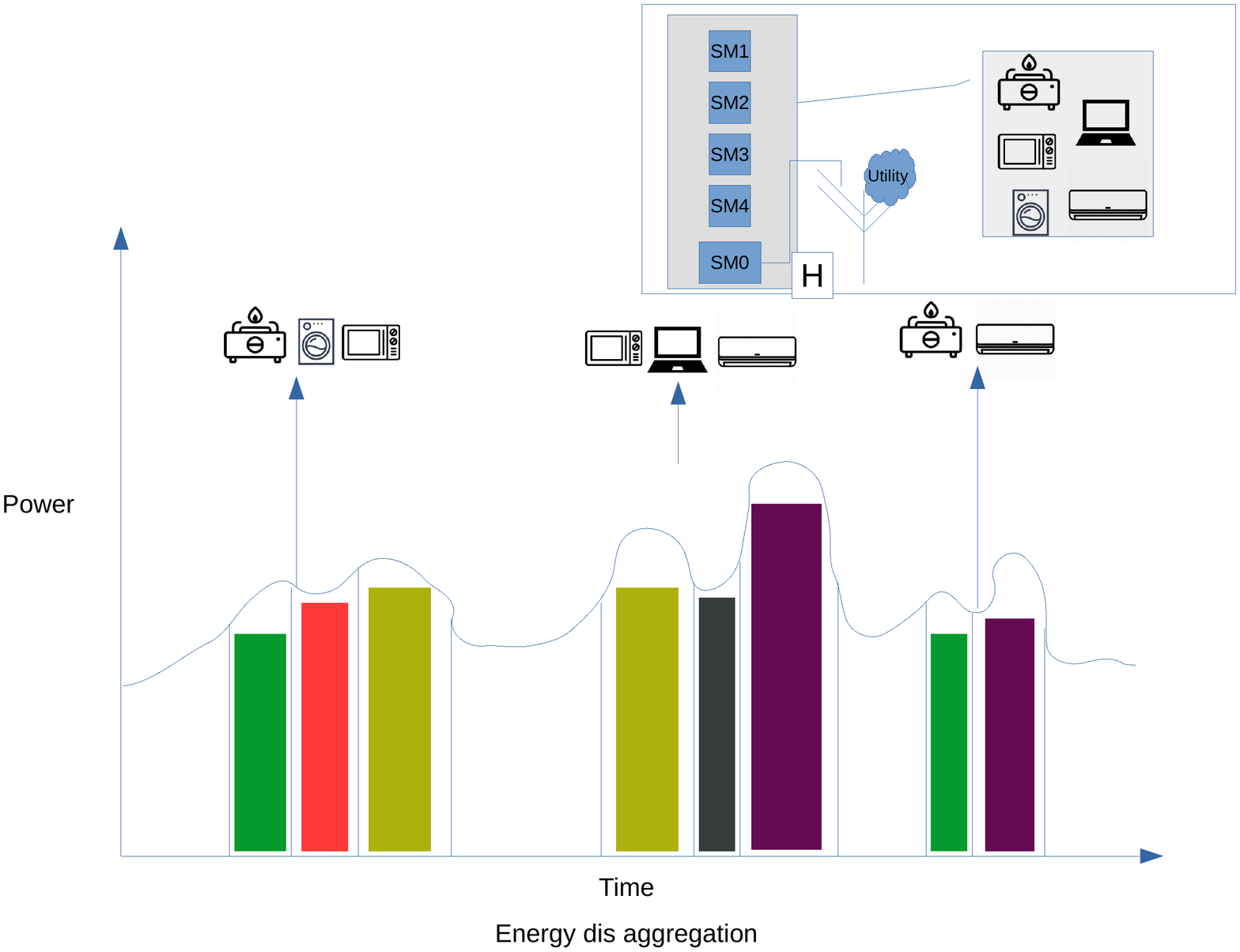}
    \centering
    \caption{Energy dis aggregation}
    \label{Fig: EN}
\end{figure}
Figure-\ref{Fig: EN} demonstrates one such scenario where a high-rise building H contains multiple apartments which are all equipped with a smart meter (SMX, where X is the apartment number) and the building itself has a smart-meter aggregator (SM0) that is connected to the electricity service. This building can be considered as a meter-group. If the electricity consumption data for this building is collected and sampled at a given rate, then applying NILM to this meter group will give us the appliance-wise energy consumption patterns for this group.

\section{State-of-the-art Dis aggregation Techniques}
\begin{figure}[]
\centering
  \includegraphics[width=0.6\columnwidth,keepaspectratio]{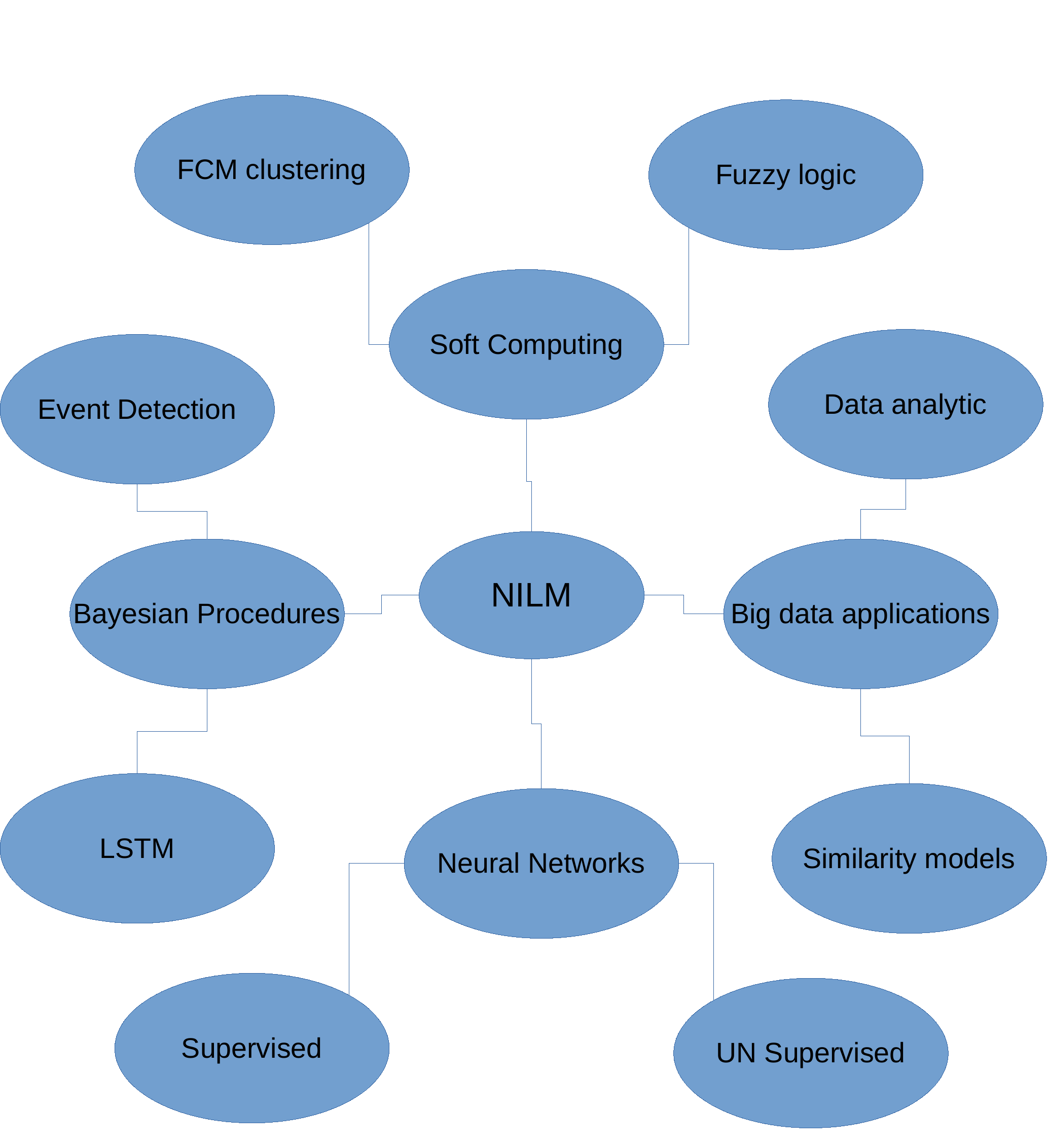}
     \centering 
     \caption{State-of-the-art Energy dis aggregation techniques}
     \label{Fig: TAX}
 \end{figure}

The method for the disaggregation of electrical measurements was proposed by Hart \cite{192069} at the beginning of the 1980s. Owing to the efforts towards energy conservation and emission reduction, there has been extensive research in this field which also got boosted due to the rise in deployment of smart-meters across the globe. There are more than 10 publicly available data sets from across different geographies to support significant research activities. The growing interest in this field has led researchers to try and implement various algorithms for solving the energy disaggregation problem, involving neural networks, big data, soft-computing and statistical methods as shown in Figure-\ref{Fig: TAX}. 

Probabilistic procedures have been gaining popularity for energy consumption modelling using a hidden Markov model (HMM)   ~\cite{kolter2011redd},   ~\cite{parson2012models},   ~\cite{zhong2015latent},   ~\cite{zhong2014signal}. The problem has been approached by implementing both supervised and unsupervised methodologies including some, where signal processing methods were adopted to enable the use of appliance's features for the purpose of disaggregation   ~\cite{7539273},   ~\cite{jin2011time}. Soft-computing techniques have also been utilized for the purpose of solving NILM by employing fuzzy clustering   ~\cite{puente2020non},  ~\cite{lin2011applications} and   ~\cite{shenavar2007novel}. Additionally, factorization procedures that aim to leverage the low-rank composition of energy consumption, also form a set of algorithms that are used to perform energy disaggregation ~\cite{batra2018transferring},\cite{batra2017matrix}, and \cite{kolter2010energy}.

Late in the year-2019, the open-source API for non-intrusive load monitoring was released as a toolkit (NILMTK). This was aimed at facilitating the means for comparisons of the different algorithms executing energy disaggregation and was intended to serve as their source library, the data set parsers needed to implement NILM, and as a reference benchmark for algorithm implementations.

The nilmtk-contrib which is the current code repository that contains the API documentation and usage examples ~\cite{batra2019towards},  lessens the amount of entry-level expertise required for algorithm developers and paves a way for algorithmic comparisons. It adds several functional enhancements to the API and eases its installation process as the disaggregation API gets rewritten. 

\subsection{Neural Networks}
Neural networks have been successfully applied to a number of load scenarios and have achieved better scores in terms of accuracy as well as producing generalization for unseen houses   ~\cite{kelly2015neural},   ~\cite{he2016empirical}. The most successfully employed neural network algorithms are Recurrent neural networks and Long Short Term Memory. There also exists an RNN based approach for NILM on small power office equipment   ~\cite{9255127}.
\subsubsection{DAE}
Autoencoders are made of simplistic neural structures and share similarities with principal component analysis.  If an autoencoder is composed of a linear activation within each of its layers,  it is highly likely that the latent variables present at the bottleneck of the encoder network (the smallest layer in the network) might directly resemble to the principal components from PCA. They basically are a set of unsupervised machine learning concepts that result in reducing the inherent dimensionality of the projected data while preserving its essential features and enable the data from a higher dimension to be synthesized in a lower dimension using non-linear transformation. It comprises of two functions, specifically, encoder and decoder. The process of data compression and down-sampling the input-data into a fewer number of bits and mapping it back to the latent space is performed by the encoder. The decoder function then puts the input encoding to use, reconstructing the input and mapping it to the latent space. The latent space is present at the bottleneck. A special deep neural network design implemented to extract a particular input component from the received noisy data is termed a denoising autoencoder. As the name suggests, it subtracts the noise from a given input in order to extract meaningful data. It forces the network to undergo overfitting on the random noise by adding some amount of white noise to the input. The error incurred is then compared to produce the required output. Removal of grain from input images and the reverb of voice signals are some popular applications of DAE. For its usage in NILM, the mains signal has been treated as a noisy appliance-power signal and the mains signal is considered to be the summation of the total power consumed by the target appliance and some additional noise. Since DAE denoises one appliance at a time, it needs multiple trained models for a group of appliances to be disaggregated. Moreover, in this scenario, the DAE gets as input a window of the mains readings (length of time-window remains fixed) and produces the induced appliance consumption values for the same time-window as output. The architecture of the network remains as was proposed in nilmtk-contrib \cite{10.1145/3360322.3360844}. 

\subsubsection{RNN}
 Recurrent Neural Networks are a set of algorithms commonly used for time series prediction, natural language processing, and other sequence processing queries. It tackles the input elements one by one and retains all the previously encountered states. The algorithm works by saving the output of a state and feeding it back to the input layer which carries it for later predictions. Therefore, every element of the model takes into consideration not only the current inputs but also what it remembers from the preceding elements. Retaining the memory of previous states gives it the ability to learn a long-term dependency by processing a series of events and manipulate the entire context at once in order to make a prediction. 
An RNN is similar to a chain-like recurrent module of a neural network but also carries a layer of memory cells. Holding onto information throughout the network leaves the RNNs with a few disadvantages, as they bring in the problem of vanishing gradients and exploding gradients. Since the gradients are the basic carriers of information in the RNN, they can sometimes either accumulate excessive information which may even contain error, or can become extremely small making the parameter updates insignificant. A very high parameter update to the neural model might weigh it down unnecessarily and remarkably small ones might make it ineffective. The resolution of this challenge is accomplished by using memory cells which can both, save a state and keep a carry to minimize information loss during sequence processing. These cells are LSTMs (Long Short Term Memory). It also has a chain-like structure but with multiple communicating neural network layers which decide the amount of data that is required to be retained, the significance of the data to be remembered and the part of the memory cell that impacts the output at the given time-step. Moreover, RNNs work well for sequential data as in the case of NILM, because they allow neuron connections in the same layer of the neural network. The employed RNN model is fed a sequence of mains readings as input and produces a single value output gives is the power consumed by the appliance at test. The network also utilizes some units of long short-term memory (LSTM) and stores values in the built-in memory cells to overcome the problem of vanishing gradients. The network architecture remains as was proposed in nilmtk-contrib  \cite{10.1145/3360322.3360844}.

\subsubsection{Sequence-to-Sequence}
The sequence to sequence learning model is a deep learning concept that is used to convert from one sequence to another. It contains an encoder RNN to understand the input sequence and a decoder RNN to decode the thought vector thereby constructing an output sequence. The process of condensing the input into a vector is achieved by an encoder network which passes its output to the decoder network where the vector is unfolded to form a new sequence. There's a set of RNNs that operate commonly between the encoder and the decoder in a sequence-to-sequence model. At each step of recurrence in the encoder network, a new word is fed to the input which gets utilized in the subsequent step by the next state. Once the decoder receives the final state from the encoder, applies a discrete probability distribution (to the input at each step) to predict the output taking into consideration, a loss function. The model here tries to map the sequence of mains reading to the target appliance sequence by learning a regressive map between them. The seq2seq model uses a regression expression which is defined by $x_t:t+W−1 = f (Y_t:t+W−1, Q)+ E_t$ where $f$ is a neural network. The architecture of the network and other hyperparameters remain as they were proposed in nilmtk-contrib \cite{10.1145/3360322.3360844}. 

\subsubsection{Sequence-to-Point}
Sequence to point learning (seq2point) operates by modifying the received network input to work as a mains window, while the output for the target appliance shows up functioning at the midpoint of its window.   
The model believes that the midpoint of the target appliance is meant to be correlated with the received mains signal information, both before and after the point of time when it occurs. This training technique can also be considered to be a non-linear regression. The architecture of the network and other hyperparameters remain as they were proposed in nilmtk-contrib \cite{10.1145/3360322.3360844}.

\subsubsection{OnlineGRU}
Gated Recurrent Unit (GRU) is a new generation of Neural Networks that replace the LSTM units with lightweight Gated Recurrent Units (GRU) thus lessening the computational requirements while delivering the equivalent performance. They also attempt to reduce redundancy by optimising the recurrent layer sizes, as well as minimising the risk of a vanishing gradient. GRUs are equipped with an internal mechanism called gates which are capable of regulating the flow of information and provide an effective solution to short-term memory. These gates work by passing information on to the long chain of sequences as they learn which part of the data is important. GRUs are similar to LSTMs except that they do not have a cell state and instead, use the hidden state for information transfer. They usually contain two gates, one for reset and update each. The reset gate determines how much of the past knowledge to forget whereas the update gate decides what part of the information to be thrown away and what new information to be added to the unit. If the reset gate approaches zero, the hidden state is forced to disregard its state and gets reset with the current input. This permits the hidden state to discard any information that's found to be insignificant within the future. This result permits a more compact representation. Whereas the upgrade gate controls how much information from the past hidden state will be exchanged to the current hidden state. The actuation of the GRU at a specific time may be a straight addition between the past actuation and the candidate activation, where an updated door chooses how much the unit overhauls its activation or content. The Online GRU model for NILM, receives as input, the latest available mains readings of the test appliance $Y_t:t+W −1$  and outputs the calculated power consumption $x_j(t+W −1)$ of the appliance at test $j$, for the last point in time. The architecture of the network and other hyperparameters remain as they were proposed in nilmtk-contrib \cite{10.1145/3360322.3360844}. 

\subsection{Big Data}
A good deal of big-data approaches has popped for energy disaggregation with the application of data analytics in smart meters   ~\cite{zhang2018big} such as Neighbourhood NILM which works on the intuition that ‘similar’ homes have ‘similar’ energy consumption on a per-appliance basis   ~\cite{batra2015neighbourhood} and a three-level learning framework that would provide different levels of knowledge abstractions for smart cities. It is scalable and also suits the nature of data generated by smart cities which can become hierarchical ~\cite{mohammadi2018enabling}.

\subsection{Soft Computing}
Non-intrusive load monitoring for residential electrical consumption has also seen success in feature extraction and pattern recognition using fuzzy logic. A set of soft-computing routines are being applied to learn and identify the power usage patterns of different devices from aggregated consumption records   ~\cite{puente2020non}.  A fuzzy classifier has been proposed which can recognise the distinctive appliance-energy states, such as energizing and de-energizing with the use of Fuzzy C-Means (FCM) clustering and optimization  ~\cite{lin2011applications}. 
\subsection{Bayesian Procedures}
Bayesian algorithms are being used for event detection and classifying the load data into the individual contribution of appliances. Findings for event detection method based on cepstrum smoothing   ~\cite{de2017bayesian} and LSTM models   ~\cite{jones2020stop} along
with a set of modules to address the issues of multi-dimensionality when the number of appliances increase   ~\cite{kaselimi2019bayesian} have been devised.

\section{Data sets and Algorithms}
The following section elaborates on the data sets that have been used for the purpose of this research experiment and the algorithms that were implemented using the NILM-TK API. The API was implemented using five neural network algorithms on three publicly available data sets.
\begin{figure*}[!htbp]
    \includegraphics[width=\textwidth,height=\textheight,keepaspectratio]{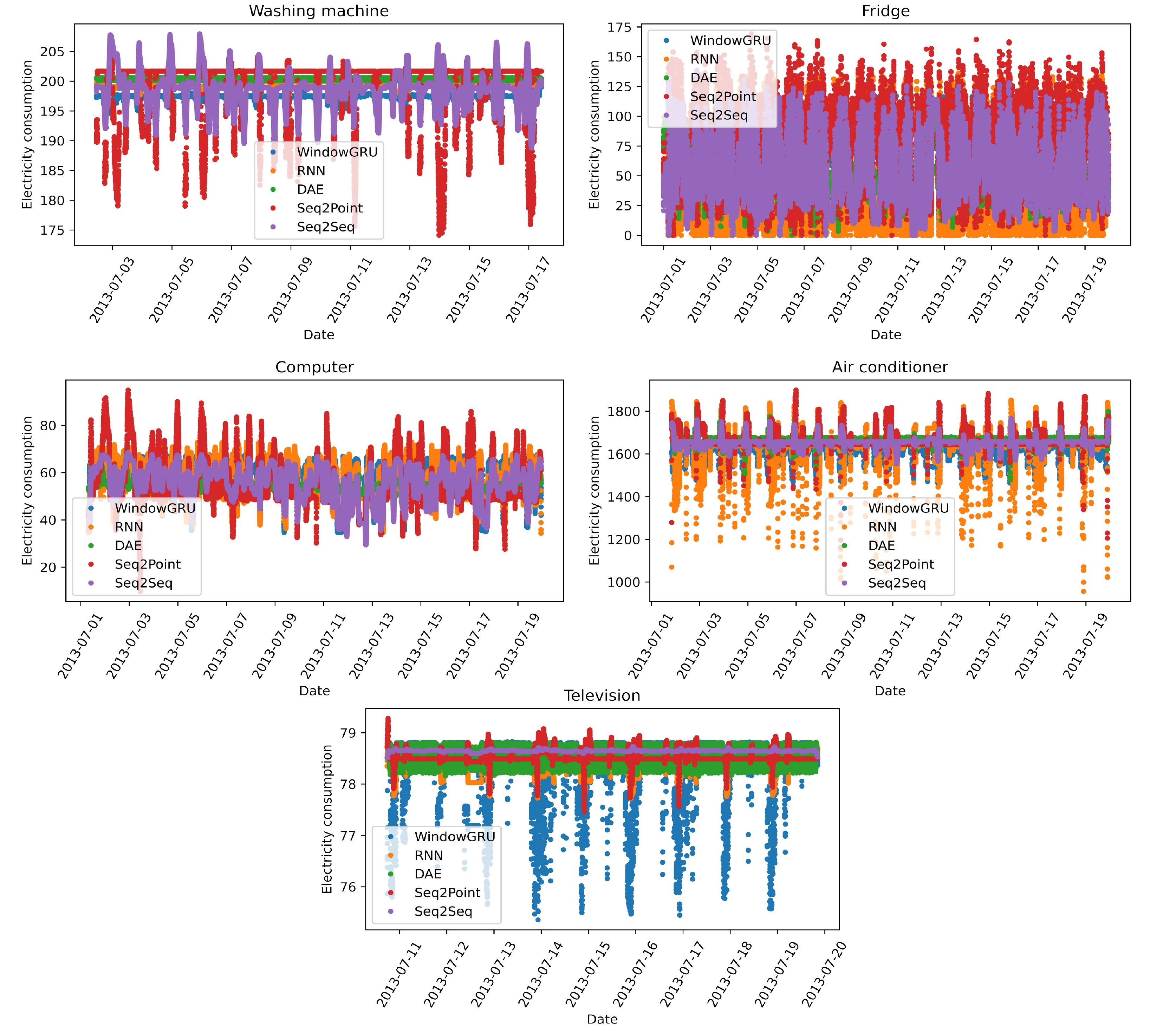}\hfill
    \caption{Predicted energy consumption: IAWE}
    \label{Fig: IAWE}
\end{figure*}
\subsection{Data sets Considered}
This analysis uses three data sets, namely REDD, UK-DALE and IAWE. The data sets used for this experiment have been taken from open-source locations.

\subsubsection{REDD}The Reference Energy Disaggregation Data Set (REDD), is a freely available data set containing detailed power usage information from several homes, which has been published with the aim of promoting further research on energy disaggregation. REDD was the first public energy dataset that was released by MIT in 2011. REDD contains high and low-frequency energy readings from 6 households in the USA recorded for short period (between a few weeks and a few months). This data set is widely used for the evaluation of NILM algorithms.

\subsubsection{UK-DALE} UK-DALE is an open-access data set from the UK, that records Domestic Appliance-Level Electricity at a sample rate of $16 kHz$ for the whole-house and at $1/6 Hz$ for individual appliances. The data set contains 16 kHz current and voltage aggregate meter readings and 6 second sub-metered power data from individual appliances across 3 UK homes, as well as 1 second aggregate and 6 second sub-metered power data for 2 additional homes. An update to this data set was released in August 2015 which has expanded the data available for house 1 to 2.5 years. The updated data set has been utilized for this experiment.
\begin{table*}[!htbp]
\centering
\caption{MAE against Ground Truth for Iawe}
\begin{tabular}{cccccc}
\hline
\textbf{MAE- Iawe} & \textbf{WindowGRU} & \textbf{RNN} & \textbf{DAE} & \textbf{Seq2Point} & \textbf{Seq2Seq} \\ \hline
washing machine    & 66.726196          & 65.665489    & 65.275597    & 65.736549          & 65.298805        \\ \hline
fridge             & 53.850956          & 68.835884    & 51.917274    & 36.775997          & 54.116047        \\ \hline
computer           & 24.112394          & 24.148933    & 24.46452     & 24.64801           & 24.25812         \\ \hline
air conditioner    & 169.634262         & 139.516449   & 139.706543   & 120.483185         & 141.557159       \\ \hline
television         & 1.347255           & 1.353503     & 1.347071     & 1.475708           & 1.449378         \\ \hline
\label{Table: IAWE}
\end{tabular}
\end{table*}
\begin{table*}[!htbp]
\centering
\caption{MAE against Ground Truth for REDD}
\begin{tabular}{cccccc}
\hline
\textbf{MAE- Redd} & \textbf{WindowGRU} & \textbf{RNN} & \textbf{DAE} & \textbf{Seq2Point} & \textbf{Seq2Seq} \\ \hline
washing machine    & 49.6175            & 60.661507    & 49.541073    & 47.514114          & 52.302326        \\ \hline
fridge             & 59.622471          & 68.384491    & 63.540127    & 55.803658          & 63.520336        \\ \hline
light              & 30.128523          & 32.435764    & 31.222496    & 31.245703          & 31.557386        \\ \hline
sockets            & 0.870423           & 0.888707     & 0.939906     & 0.827897           & 0.883221         \\ \hline
microwave          & 28.598331          & 28.599768    & 34.998123    & 21.999201          & 24.803219        \\ \hline
\label{Table: REDD}
\end{tabular}
\end{table*}

\begin{table*}[!htbp]
\centering
\caption{MAE against Ground Truth for UKdale}
\begin{tabular}{cccccc}
\hline
\textbf{MAE-UKdale} & \textbf{WindowGRU} & \textbf{RNN} & \textbf{DAE} & \textbf{Seq2Point} & \textbf{Seq2Seq} \\ \hline
washing machine      & 50.239201          & 55.136917    & 53.854832    & 49.03191           & 43.448345        \\ \hline
fridge               & 44.169342          & 44.513477    & 45.257683    & 44.960304          & 44.8424          \\ \hline
light                & 33.446453          & 34.21154     & 35.531647    & 33.719086          & 35.57534         \\ \hline
dish washer          & 49.290272          & 47.874489    & 39.505829    & 49.656605          & 41.864983        \\ \hline
microwave            & 21.169703          & 16.606073    & 25.805212    & 20.247181          & 19.278296        \\ \hline
\label{Table: UKdale}
\end{tabular}

\end{table*}
\subsubsection{IAWE} The IAWE data set was released by the Indraprastha Institute of Information Technology, India. This data contains the aggregated and sub-metered electricity and gas readings from 33 household sensors, captured at a resolution of one second. The data set covers 73 days of a single house in Delhi, India.

\subsection{Algorithms Considered}
For the purpose of this experiment, five neural network based algorithms have been utilised, namely: DAE, RNN, Sequence-to-Sequence, Sequence-to-Point and OnlineGRU, all of which have been described in detail in section 4.1 of this article.

\begin{figure*}[!htbp]
    \includegraphics[width=\textwidth,height=\textheight,keepaspectratio]{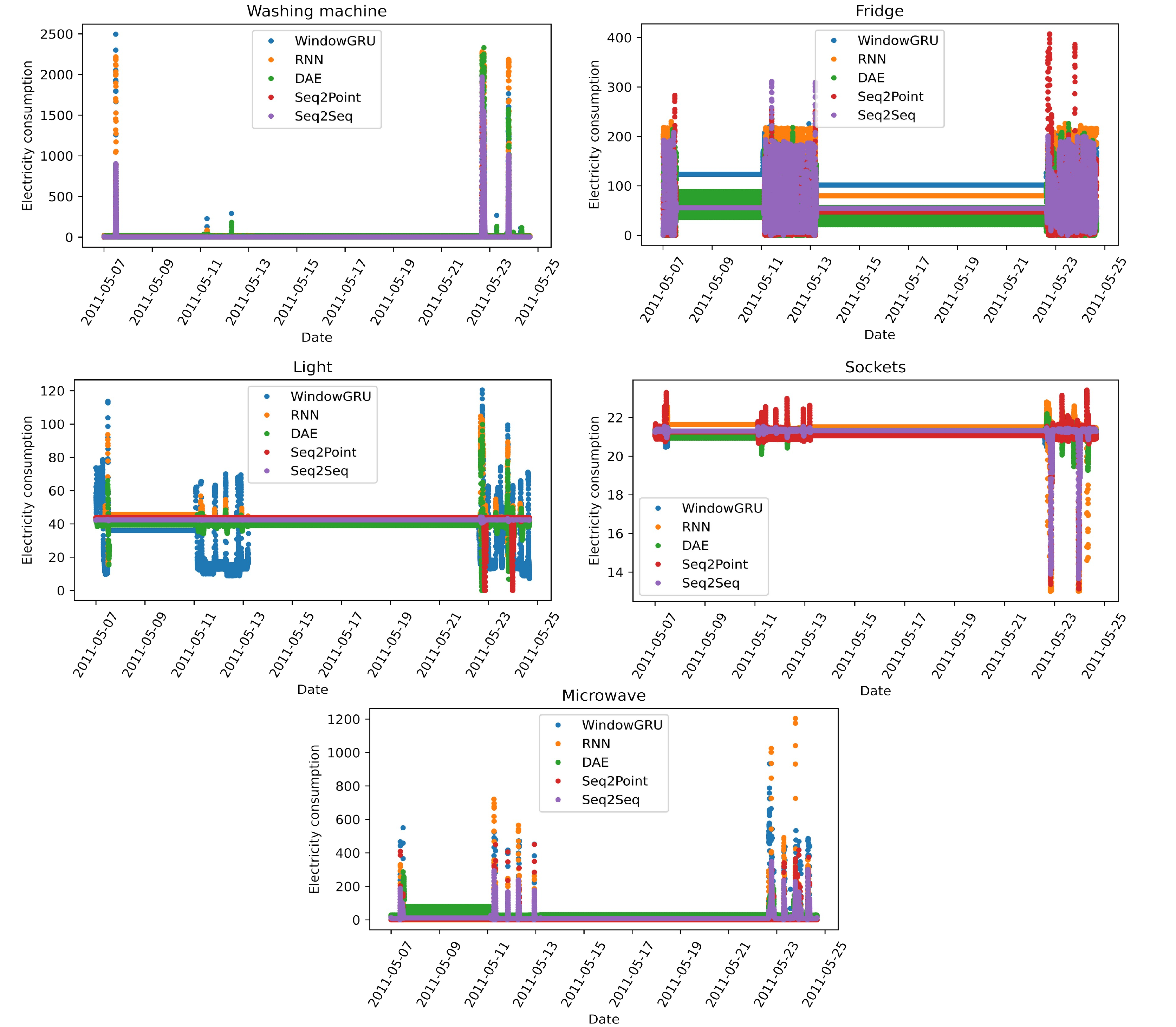}\hfill
    \caption{Predicted energy consumption: REDD}
    \label{Fig: REDD}
\end{figure*}

\subsection{Experimental Scenario} 
The tests were run on a machine with GeForce GTX 1660 Ti/PCIe/SSE2 GPUs with Intel® Core™ i7-9750H CPU @ 2.60GHz × 12 and 16 GB RAM.  The batch size was 32, for all neural algorithms and they were trained for 5 epochs each. The sample period was kept as 60 seconds.

\subsection{Results and Discussion} For each data set, the network was trained for a period of 30 days based on the data and tested on the subsequent 20 days using the nilmtk-contrib API to predict the energy consumed per appliance and report the mean absolute error(MAE) in each case. The outputs from the tests on the three datasets under consideration do not collectively reveal the best performer out of the five algorithms used. Moreover, there is no clear winner in each case except UK-dale where Sequence-to-Sequence leads. This could be due to the differences in the data used.
Mean absolute error calculates the total magnitude of error in a set of predictions, not considering their direction. 
It assumes that all individual differences are supposed to have equal weight and therefore, for a test sample, it gives out the absolute differences between the actual and predicted observation. 
It can be used to represent the average prediction error in terms of the variable at the test. The metric ranges from 0 to $\infty$ and is unresponsive to the direction of errors. This score has a negative orientation, the lower the value, the better. It effectively describes the magnitude of residuals and does not indicate the underperformance or overperformance of the model. All elements are given equal weightage and hence contribute directly to the total value of the error and larger the value, linear is the contribution.
\begin{figure*}[!htbp]
    \includegraphics[width=\textwidth,height=\textheight,keepaspectratio]{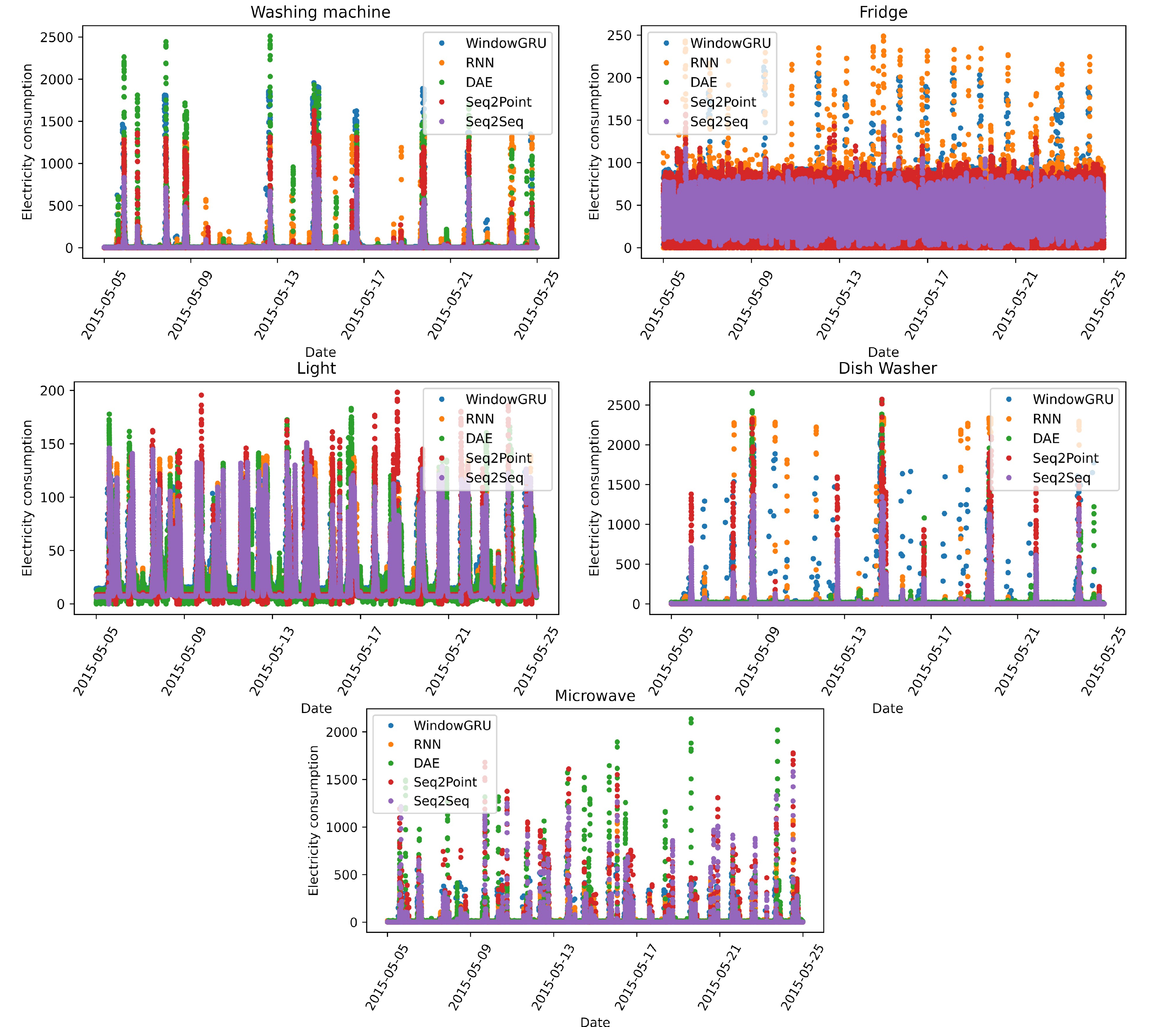}\hfill
    \caption{Predicted energy consumption: UKdale}
    \label{Fig: UKdale}
\end{figure*}
It is noticeable from Table-\ref{Table: IAWE} that there is no clear winner in the case of iAWE and the mean absolute errors are very high which could be because the data set is based on a single housing facility. The mean absolute errors remain similar in cases of washing machine, computer and television with the reported errors being the minimum for television. The case of air conditioner reports the maximum observed error from all three scenarios which is greater than 100 for all five algorithms and varies largely for every algorithm. The performance of Sequence-to-Sequence and DAE remains similar across all five appliances with DAE performing the best for the television and Sequence-to-Sequence performing the best for the washing machine. It can also be observed from Figure-\ref{Fig: IAWE}, the difference in the load cycles of different appliances.

It is evident from Table-\ref{Table: REDD} that none of the algorithms in consideration perform the best overall in the case of REDD although Sequence-to-Point leads in four of the five most used appliances. This could be due to the dissimilar usage trends of the appliances that the graphs clearly demonstrate in Figure-\ref{Fig: REDD}, hence, signifying the energy consumption patterns of the appliances. The mean absolute errors are the lowest in this case making it a preferred data set to base our conclusions on.

The mean absolute errors in for the sockets remain less than 1 which is the minimum error reported by an algorithm in all three scenarios. The errors show maximum variation in the case of the washing machine followed by the fridge. There is little variation in the reported values of mean absolute error for the sockets which is again followed by the light. The performance of WindowGRU and RNN is comparable across light, sockets and microwave but differ largely in cases of the washing machine and fridge.	

Table-\ref{Table: UKdale} shows that Sequence-to-Sequence performs the best for most cases of UK-Dale and the mean absolute errors, although high, are similar for nearly all appliances. The predictions from the graph indicate how some appliances have clearly defined usage patterns (as seen in Figure-\ref{Fig: UKdale}) whereas others are used on a daily basis. The mean absolute error stays the lowest in the case of microwave with 16.60 being the minimum as reported by RNN. It can also be noticed that the mean absolute error for the fridge remains similar across all five algorithms followed by little variation in the case of light and shows the maximum variation for the washing machine.

Different appliances use different quantities of energy and, thus, compared to a high energy consumption load, the errors measured relative to low energy use might be less significant. Furthermore, some end devices operate less frequently than others, so if a statistically substantial number of run times is not recorded, metric results may not be indicative of efficiency. In view of these problems, it is recommended that every metric assessment reflects some leveling of the use of energy or other basis for appropriate comparisons of results across different end uses which coule be a fixed energy usage per end use, a fixed number of real events per end use, or a fixed period that is capable of capturing a variety of conditions.

\section{Conclusion and future work}In this paper, we demonstrated how nilmtk-contrib provides an interface to energy disaggregation problems and also the advantages of the NILM API which enables comparison of algorithms in this domain even without expertise in the field, thus enabling efficient experimental evaluations. Consequently, acing development within the field and supporting the progress of research in NILM.

Our experimental results carried out on three publicly available data sets namely IAWE, REDD and UKdale do not indicate the presence of any patterns in the output. 
This could be a result of the large differences in the data sets. Although, it suggests that the API can be used for data sets from different geographies.

With the need for energy and resources rising each day, careful and efficient utilisation is the only way, to conserve it and energy dis-aggregation can be an essential element in the conservation of energy since it elaborates the energy usage tendencies. 
The trends observed in the energy-usage patterns from a household can be used for the purpose of security as an anomaly in them might represent a sign of appliance failure or illegal use of supplied electricity. The appliance usage patterns can also be used to calculate and control the amount of carbon emissions. 

\bibliographystyle{IEEEtran}

\bibliography{cas-refs}

\begin{thebibliography}{10}
\providecommand{\url}[1]{#1}
\csname url@samestyle\endcsname
\providecommand{\newblock}{\relax}
\providecommand{\bibinfo}[2]{#2}
\providecommand{\BIBentrySTDinterwordspacing}{\spaceskip=0pt\relax}
\providecommand{\BIBentryALTinterwordstretchfactor}{4}
\providecommand{\BIBentryALTinterwordspacing}{\spaceskip=\fontdimen2\font plus
\BIBentryALTinterwordstretchfactor\fontdimen3\font minus
  \fontdimen4\font\relax}
\providecommand{\BIBforeignlanguage}[2]{{%
\expandafter\ifx\csname l@#1\endcsname\relax
\typeout{** WARNING: IEEEtran.bst: No hyphenation pattern has been}%
\typeout{** loaded for the language `#1'. Using the pattern for}%
\typeout{** the default language instead.}%
\else
\language=\csname l@#1\endcsname
\fi
#2}}
\providecommand{\BIBdecl}{\relax}
\BIBdecl

\bibitem{shorrock2003domestic}
L.~Shorrock, J.~Utley \emph{et~al.}, \emph{Domestic energy fact file
  2003}.\hskip 1em plus 0.5em minus 0.4em\relax Citeseer, 2003.

\bibitem{darby2006effectiveness}
S.~Darby \emph{et~al.}, ``The effectiveness of feedback on energy
  consumption,'' \emph{A Review for DEFRA of the Literature on Metering,
  Billing and direct Displays}, vol. 486, no. 2006, p.~26, 2006.

\bibitem{192069}
G.~W. {Hart}, ``Nonintrusive appliance load monitoring,'' \emph{Proceedings of
  the IEEE}, vol.~80, no.~12, pp. 1870--1891, 1992.

\bibitem{zoha2012non}
A.~Zoha, A.~Gluhak, M.~A. Imran, and S.~Rajasegarar, ``Non-intrusive load
  monitoring approaches for disaggregated energy sensing: A survey,''
  \emph{Sensors}, vol.~12, no.~12, pp. 16\,838--16\,866, 2012.

\bibitem{10.1145/3360322.3360844}
\BIBentryALTinterwordspacing
N.~Batra, R.~Kukunuri, A.~Pandey, R.~Malakar, R.~Kumar, O.~Krystalakos,
  M.~Zhong, P.~Meira, and O.~Parson, ``Towards reproducible state-of-the-art
  energy disaggregation,'' in \emph{Proceedings of the 6th ACM International
  Conference on Systems for Energy-Efficient Buildings, Cities, and
  Transportation}, ser. BuildSys '19.\hskip 1em plus 0.5em minus 0.4em\relax
  New York, NY, USA: Association for Computing Machinery, 2019, p. 193–202.
  [Online]. Available: \url{https://doi.org/10.1145/3360322.3360844}
\BIBentrySTDinterwordspacing

\bibitem{8861646}
M.~{Sun}, F.~M. {Nakoty}, Q.~{Liu}, X.~{Liu}, Y.~{Yang}, and T.~{Shen},
  ``Non-intrusive load monitoring system framework and load disaggregation
  algorithms: A survey,'' in \emph{2019 International Conference on Advanced
  Mechatronic Systems (ICAMechS)}, 2019, pp. 284--288.

\bibitem{faustine2017survey}
A.~Faustine, N.~H. Mvungi, S.~Kaijage, and K.~Michael, ``A survey on
  non-intrusive load monitoring methodies and techniques for energy
  disaggregation problem,'' \emph{arXiv preprint arXiv:1703.00785}, 2017.

\bibitem{HIMEUR2020110404}
\BIBentryALTinterwordspacing
Y.~Himeur, A.~Alsalemi, F.~Bensaali, and A.~Amira, ``Building power consumption
  datasets: Survey, taxonomy and future directions,'' \emph{Energy and
  Buildings}, vol. 227, p. 110404, 2020. [Online]. Available:
  \url{http://www.sciencedirect.com/science/article/pii/S037877882030815X}
\BIBentrySTDinterwordspacing

\bibitem{klemenjak2018performance}
C.~Klemenjak, ``On performance evaluation and machine learning approaches in
  non-intrusive load monitoring,'' \emph{Energy Informatics}, vol.~1, no.~1,
  pp. 391--395, 2018.

\bibitem{haq2018prospects}
A.~U. Haq and H.-A. Jacobsen, ``Prospects of appliance-level load monitoring in
  off-the-shelf energy monitors: A technical review,'' \emph{Energies},
  vol.~11, no.~1, p. 189, 2018.

\bibitem{zhuang2018overview}
M.~Zhuang, M.~Shahidehpour, and Z.~Li, ``An overview of non-intrusive load
  monitoring: Approaches, business applications, and challenges,'' in
  \emph{2018 International Conference on Power System Technology
  (POWERCON)}.\hskip 1em plus 0.5em minus 0.4em\relax IEEE, 2018, pp.
  4291--4299.

\bibitem{jiang2011literature}
L.~Jiang, J.~Li, S.~Luo, J.~Jin, and S.~West, ``Literature review of power
  disaggregation,'' in \emph{Proceedings of 2011 International Conference on
  Modelling, Identification and Control}.\hskip 1em plus 0.5em minus
  0.4em\relax IEEE, 2011, pp. 38--42.

\bibitem{mayhorn2016load}
E.~T. Mayhorn, G.~P. Sullivan, J.~M. Petersen, R.~S. Butner, and E.~M. Johnson,
  ``Load disaggregation technologies: real world and laboratory performance,''
  \emph{Pacific Northwest National Laboratory (PNNL), Richland, WA (US), Tech.
  Rep. PNNL-SA-116560}, 2016.

\bibitem{pereira2018performance}
L.~Pereira and N.~Nunes, ``Performance evaluation in non-intrusive load
  monitoring: Datasets, metrics, and tools—a review,'' \emph{Wiley
  Interdisciplinary Reviews: data mining and knowledge discovery}, vol.~8,
  no.~6, p. e1265, 2018.

\bibitem{herrero2017non}
J.~R. Herrero, {\'A}.~L. Murciego, A.~L. Barriuso, D.~H. de~La~Iglesia, G.~V.
  Gonz{\'a}lez, J.~M.~C. Rodr{\'\i}guez, and R.~Carreira, ``Non intrusive load
  monitoring (nilm): A state of the art,'' in \emph{International Conference on
  Practical Applications of Agents and Multi-Agent Systems}.\hskip 1em plus
  0.5em minus 0.4em\relax Springer, 2017, pp. 125--138.

\bibitem{aladesanmi2015overview}
E.~Aladesanmi and K.~Folly, ``Overview of non-intrusive load monitoring and
  identification techniques,'' \emph{IFAC-PapersOnLine}, vol.~48, no.~30, pp.
  415--420, 2015.

\bibitem{6486814}
M.~{Zeifman}, K.~{Roth}, and J.~{Stefan}, ``Automatic recognition of major
  end-uses in disaggregation of home energy display data,'' in \emph{2013 IEEE
  International Conference on Consumer Electronics (ICCE)}, 2013, pp. 104--105.

\bibitem{10.1145/2602044.2602051}
\BIBentryALTinterwordspacing
N.~Batra, J.~Kelly, O.~Parson, H.~Dutta, W.~Knottenbelt, A.~Rogers, A.~Singh,
  and M.~Srivastava, ``Nilmtk: An open source toolkit for non-intrusive load
  monitoring,'' in \emph{Proceedings of the 5th International Conference on
  Future Energy Systems}, ser. e-Energy '14.\hskip 1em plus 0.5em minus
  0.4em\relax New York, NY, USA: Association for Computing Machinery, 2014, p.
  265–276. [Online]. Available: \url{https://doi.org/10.1145/2602044.2602051}
\BIBentrySTDinterwordspacing

\bibitem{kolter2011redd}
J.~Z. Kolter and M.~J. Johnson, ``Redd: A public data set for energy
  disaggregation research,'' in \emph{Workshop on data mining applications in
  sustainability (SIGKDD), San Diego, CA}, vol.~25, no. Citeseer, 2011, pp.
  59--62.

\bibitem{parson2012models}
O.~Parson, S.~Ghosh, M.~Weal, A.~Rogers, and N.-I. L. M.~U. Prior, ``Models of
  general appliance types,'' 2012.

\bibitem{zhong2015latent}
M.~Zhong, N.~Goddard, and C.~Sutton, ``Latent bayesian melding for integrating
  individual and population models,'' in \emph{Advances in neural information
  processing systems}, 2015, pp. 3618--3626.

\bibitem{zhong2014signal}
------, ``Signal aggregate constraints in additive factorial hmms, with
  application to energy disaggregation,'' in \emph{Advances in Neural
  Information Processing Systems}, 2014, pp. 3590--3598.

\bibitem{7539273}
K.~{He}, L.~{Stankovic}, J.~{Liao}, and V.~{Stankovic}, ``Non-intrusive load
  disaggregation using graph signal processing,'' \emph{IEEE Transactions on
  Smart Grid}, vol.~9, no.~3, pp. 1739--1747, 2018.

\bibitem{jin2011time}
Y.~Jin, E.~Tebekaemi, M.~Berges, and L.~Soibelman, ``A time-frequency approach
  for event detection in non-intrusive load monitoring,'' in \emph{Signal
  Processing, Sensor Fusion, and Target Recognition XX}, vol. 8050.\hskip 1em
  plus 0.5em minus 0.4em\relax International Society for Optics and Photonics,
  2011, p. 80501U.

\bibitem{puente2020non}
C.~Puente, R.~Palacios, Y.~Gonz{\'a}lez-Arechavala, and E.~F.
  S{\'a}nchez-{\'U}beda, ``Non-intrusive load monitoring (nilm) for energy
  disaggregation using soft computing techniques,'' \emph{Energies}, vol.~13,
  no.~12, p. 3117, 2020.

\bibitem{lin2011applications}
Y.-H. Lin, M.-S. Tsai, and C.-S. Chen, ``Applications of fuzzy classification
  with fuzzy c-means clustering and optimization strategies for load
  identification in nilm systems,'' in \emph{2011 IEEE international conference
  on fuzzy systems (FUZZ-IEEE 2011)}.\hskip 1em plus 0.5em minus 0.4em\relax
  IEEE, 2011, pp. 859--866.

\bibitem{shenavar2007novel}
P.~Shenavar and E.~Farjah, ``Novel embedded real-time nilm for electric loads
  disaggregating and diagnostic,'' in \emph{EUROCON 2007-The International
  Conference on" Computer as a Tool"}.\hskip 1em plus 0.5em minus 0.4em\relax
  IEEE, 2007, pp. 1555--1560.

\bibitem{batra2018transferring}
N.~Batra and Y.~Jia, ``Transferring decomposed tensors for scalable energy
  breakdown across regions,'' in \emph{The Thirty-Second AAAI Conference on
  Artificial Intelligence (AAAI-18)}, 2018.

\bibitem{batra2017matrix}
N.~Batra, H.~Wang, A.~Singh, and K.~Whitehouse, ``Matrix factorisation for
  scalable energy breakdown,'' in \emph{Proceedings of the... AAAI Conference
  on Artificial Intelligence}, 2017.

\bibitem{kolter2010energy}
J.~Z. Kolter, S.~Batra, and A.~Y. Ng, ``Energy disaggregation via
  discriminative sparse coding,'' in \emph{Advances in Neural Information
  Processing Systems}, 2010, pp. 1153--1161.

\bibitem{batra2019towards}
N.~Batra, R.~Kukunuri, A.~Pandey, R.~Malakar, R.~Kumar, O.~Krystalakos,
  M.~Zhong, P.~Meira, and O.~Parson, ``Towards reproducible state-of-the-art
  energy disaggregation,'' in \emph{Proceedings of the 6th ACM International
  Conference on Systems for Energy-Efficient Buildings, Cities, and
  Transportation}, 2019, pp. 193--202.

\bibitem{kelly2015neural}
J.~Kelly and W.~Knottenbelt, ``Neural nilm: Deep neural networks applied to
  energy disaggregation,'' in \emph{Proceedings of the 2nd ACM international
  conference on embedded systems for energy-efficient built environments},
  2015, pp. 55--64.

\bibitem{he2016empirical}
W.~He and Y.~Chai, ``An empirical study on energy disaggregation via deep
  learning,'' in \emph{2016 2nd International Conference on Artificial
  Intelligence and Industrial Engineering (AIIE 2016)}.\hskip 1em plus 0.5em
  minus 0.4em\relax Atlantis Press, 2016, pp. 338--342.

\bibitem{9255127}
M.~. T.~E. {Astal}, M.~{Kalloub}, A.~M. {Abu-Hudrouss}, and G.~{Frey}, ``Office
  appliances identification and monitoring using deep leaning based energy
  disaggregation for smart buildings,'' in \emph{IECON 2020 The 46th Annual
  Conference of the IEEE Industrial Electronics Society}, 2020, pp. 1986--1991.

\bibitem{zhang2018big}
Y.~Zhang, T.~Huang, and E.~F. Bompard, ``Big data analytics in smart grids: a
  review,'' \emph{Energy informatics}, vol.~1, no.~1, pp. 1--24, 2018.

\bibitem{batra2015neighbourhood}
N.~Batra, A.~Singh, and K.~Whitehouse, ``Neighbourhood nilm: A big-data
  approach to household energy disaggregation,'' \emph{arXiv preprint
  arXiv:1511.02900}, 2015.

\bibitem{mohammadi2018enabling}
M.~Mohammadi and A.~Al-Fuqaha, ``Enabling cognitive smart cities using big data
  and machine learning: Approaches and challenges,'' \emph{IEEE Communications
  Magazine}, vol.~56, no.~2, pp. 94--101, 2018.

\bibitem{de2017bayesian}
L.~De~Baets, J.~Ruyssinck, C.~Develder, T.~Dhaene, and D.~Deschrijver, ``On the
  bayesian optimization and robustness of event detection methods in nilm,''
  \emph{Energy and Buildings}, vol. 145, pp. 57--66, 2017.

\bibitem{jones2020stop}
R.~Jones, C.~Klemenjak, S.~Makonin, and I.~V. Baji{\'c}, ``Stop: Exploring
  bayesian surprise to better train nilm,'' in \emph{Proceedings of the 5th
  International Workshop on Non-Intrusive Load Monitoring}, 2020, pp. 39--43.

\bibitem{kaselimi2019bayesian}
M.~Kaselimi, N.~Doulamis, A.~Doulamis, A.~Voulodimos, and E.~Protopapadakis,
  ``Bayesian-optimized bidirectional lstm regression model for non-intrusive
  load monitoring,'' in \emph{ICASSP 2019-2019 IEEE International Conference on
  Acoustics, Speech and Signal Processing (ICASSP)}.\hskip 1em plus 0.5em minus
  0.4em\relax IEEE, 2019, pp. 2747--2751.

\end{thebibliography}

\begin{IEEEbiography}[{\includegraphics[width=1in,height=1in,clip,keepaspectratio]{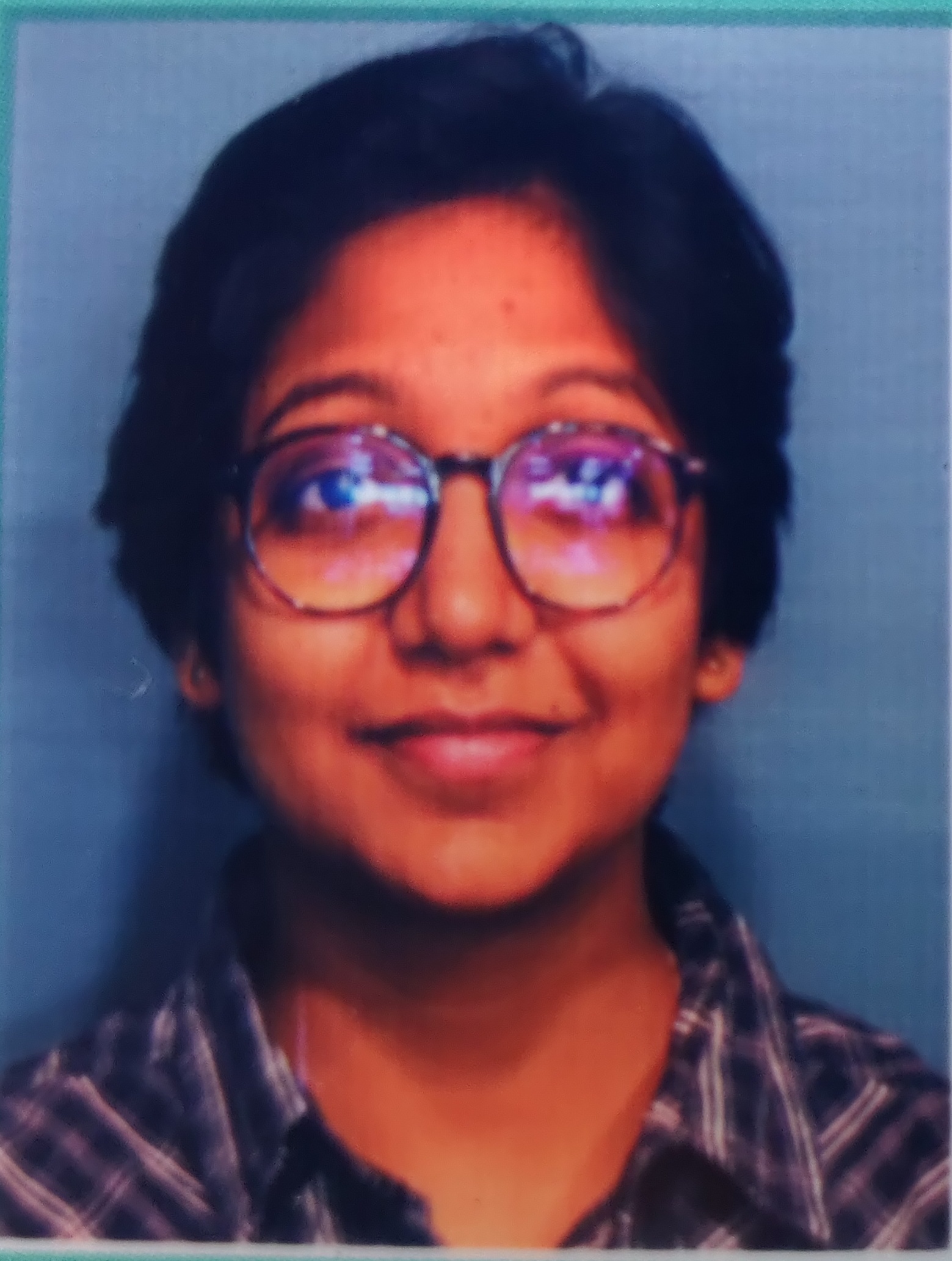}}]{Akriti Verma} is a postgraduate student studying Applied Artificial Intelligence at Deakin University, Australia. Previously she has worked as a Researcher at Tata Consultancy Services, India. She completed her graduation in Information Technology from AKTU, India. She is driven by her interest in analysis which when incorporated with AI principles enables her to understand and engineer solutions for real-life use cases. She wishes to pursue research in the same direction.
\end{IEEEbiography}
\begin{IEEEbiography}[{\includegraphics[width=1in,height=1.25in,clip,keepaspectratio]{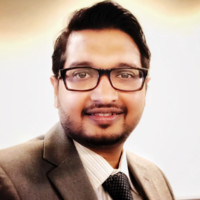}}]{Adnan Anwar} is a Lecturer and Deputy Director of postgraduate cybersecurity studies at the School of Information Technology. Previously he has worked as a Data Scientist at Flow Power. He has over 8 years of research, and teaching experience in universities and research labs including NICTA, La Trobe University, and University of New South Wales. He received his PhD and sMaster by Research degree from UNSW. He is broadly interested in the security research for critical infrastructures including Smart Energy Grid, SCADA system, and application of machine learning and optimization techniques to solve cyber security issues for industrial systems. He has been the recipient of several awards including UPA scholarship, UNSW TFR scholarship, best paper award and several travel grants including ACM and Postgraduate Research Student Support (PRSS) travel grants. He has authored over 40 articles including high-impact journals (mostly in Q1), conference articles and book chapters in prestigious venues. He is an active member of IEEE for over 9 years and serving different committees. 
\end{IEEEbiography}

\begin{IEEEbiography}[{\includegraphics[width=1in,height=1.25in,clip,keepaspectratio]{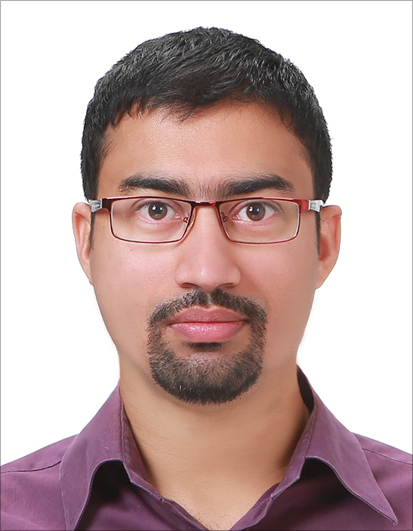}}]{M. A. PARVEZ MAHMUD}received his B.Sc. degree in Electrical and Electronic Engineering and Master of Engineering degree in Mechatronics Engineering. After the successful completion of his Ph.D. degree with multiple awards, he worked as a Postdoctoral Research Associate and Academic in the School of Engineering at Macquarie University, Sydney. He is currently an Alfred Deakin Postdoctoral Research Fellow at Deakin University. He worked at World University of Bangladesh (WUB) as a ‘Lecturer’ for more than 2 years and at the Korea Institute of Machinery and Materials (KIMM) as a ‘Researcher’ for about 3 years. His research is focused on Energy Sustainability, Secure Energy Trading, Microgrid Control and Economic Optimization, Machine Learning, Data Science, and Micro/nanoscaled Technologies for Sensing and Energy Harvesting. He accumulated experience and expertise in machine learning, life cycle assessment, sustainability and economic analysis, materials engineering, microfabrication, and nanostructured energy materials to facilitate technological translation from the lab to real-world applications for the better society.   He has produced over 50 publications, including 1 authored book, 3 Book Chapters, 29 Journal Papers, and 21 fully refereed Conference Papers. He received several awards including “Macquarie University Highly Commended Excellence in Higher Degree Research Award 2019”. He was involved in teaching engineering subjects in the Electrical, Biomedical and Bechatronics Engineering courses at the School of Engineering, Macquarie University for more than 2 years. Currently, he is involved in the supervision of 6 PhD students at Deakin University. He is a key member of Deakin University’s Advanced Integrated Microsystems (AIM) research group. Apart from this, he is actively involved with different professional organizations, including Engineers Australia and IEEE. 
\end{IEEEbiography}

\begin{IEEEbiography}[{\includegraphics[width=1in,height=1.25in,clip,keepaspectratio]{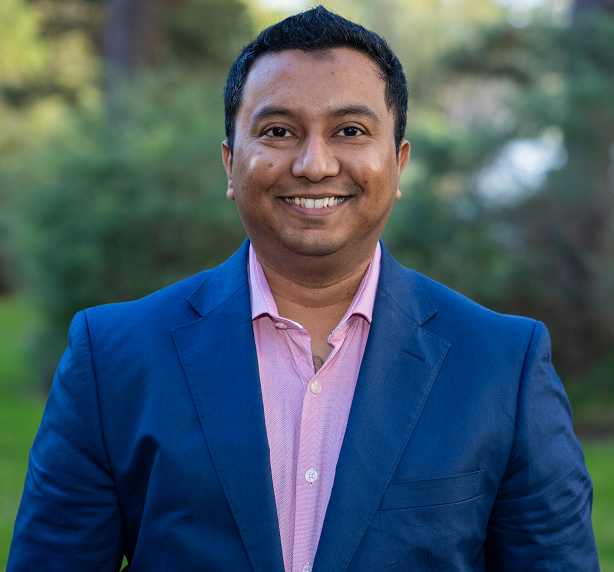}}]{Mohiuddin Ahmed} is currently working as a Lecturer with the Academic Centre for Cyber Security Excellence, Edith Cowan University, Australia. His research intersects between data analytics and cybersecurity. His research has a high impact on data and security analytics, false data injection attacks, and digital health. Mohiuddin has led edited multiple books and contributed articles in The Conversation. He has over 50 publications in reputed venues. Mohiuddin secured the prestigious ECU Early Career Researcher Grant for investigating effectiveness of blockchain for dependable and secure e-health. He also secured several external and internal grants within a very short time frame. He is a Senior Member of IEEE and member of Australian Computer Society/certified professional.
\end{IEEEbiography}

\begin{IEEEbiography}[{\includegraphics[width=1in,height=1.25in,clip,keepaspectratio]{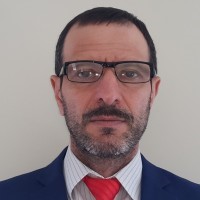}}]{Prof. Abbas Kouzani} received his B.Sc. degree in Computer Engineering from Sharif University of Technology, Iran, his M.Eng.Sc. degree is Electrical and Electronic Engineering from University of Adelaide, Australia, and his Ph.D. degree in Electrical and Electronic Engineering from Flinders University, Australia. He was a lecturer with the School of Engineering, Deakin University, and then a Senior Lecturer with the School of Electrical Engineering and Computer Science, University of Newcastle, Australia. Currently, he is a Professor with the School of Engineering, Deakin University, Australia. He provides research leadership in embedded, connected, and low-power devices, circuits, and instruments that incorporate sensing, actuation, control, wireless transmission, networking and IoT, data acquisition/storage/analysis, AI, energy harvesting, power management, and fabrication for tackling research questions relating to a variety of disciplines including healthcare, ecology, mining, infrastructure, automotive, manufacturing, energy, utilities, and agriculture. Has produced over 370 publications including 1 book, 17 Book Chapters, 180 Journal Papers, and 181 fully refereed Conference Papers. He has 3 patents and 2 pending patents. He has been involved in over \$15 million research grants, and has managed projects and delivered research solutions to over 25 Australian and International companies. He received several awards including “Outstanding Contribution to Scholarly Publication Award”, School of Engineering, Deakin University, 2019. He has supervised 24 research fellows/assistants, and produced 28 Ph.D. and 6 Masters by Research completions. Currently, he is involved in supervision of 12 PhD students. He is the Director of Deakin University’s Advanced Integrated Microsystems (AIM) research group.
\end{IEEEbiography}

\EOD

\end{document}